\documentstyle[pra,aps,multicol,psfig]{revtex}

\def\addcontentsline#1#2#3{\relax}
\begin{document}
\draft
\title{Fractional charge in transport through a 1D correlated insulator
of finite length.}
\author{V.V. Ponomarenko$^{1,2}$ and N. Nagaosa$^{1}$}
\address{$^1$ Department of Applied Physics, University of Tokyo,
Bunkyo-ku, Tokyo 113, Japan\\
$^2$ A.F.Ioffe Physical Technical Institute,
194021, St. Petersburg, Russia}
\date{\today}
\maketitle
\begin{abstract}
Transport through a one channel wire of length $L$
confined between two leads
is examined when the 1D electron system has an energy
gap $2M$: $M > T_L \equiv v_c/L$ induced by the interaction in charge
mode ($v_c$: charge velocity in the wire). In spinless case 
the transformation of
the leads electrons into the charge density wave
solitons of fractional charge $q$ entails a non-trivial low energy
crossover from the Fermi liquid behavior below 
the crossover energy $T_x \propto \sqrt{T_L M}
e^{-M /[T_L(1-q^2)]}$ to
the insulator one with the fractional charge in current vs. voltage, 
conductance vs. temperature, and in shot noise. Similar behavior is 
predicted
for the Mott insulator of filling factor $\nu = integer/(2 m')$.

\end{abstract}


%\narrowtext
%
\multicols{2}
Fractional charge (FC) is one of the central issues in the physics of
strongly correlated electronic system. Recent measurements of the
shot noise \cite{expnoise} have shown the existence of the quasi-particles
with charge $q= \pm e/3$ in fractional quantum Hall liquid (FQHL) at
filling $\nu =1/3$. In this incompressible system the quantum 
transport is dominated by the edge mode which has been described 
as a chiral Tomonaga-Luttinger liquid (TLL). Another realization
of TLL is the quantum wire made out of $GaAs$. Recently Tarucha {\it et al. }
succeeded to introduce a periodic potential to it, which 
allows Umklapp scattering and, hence, the insulating behavior due to
correlation. For the spinless electrons such a 1D insulator is also
a charge density wave insulator (CDW). Transport through it reduces to mutual transformations between the reservoir electrons 
and the CDW (anti)solitons of FC $q \le e \equiv 1$
and mass $M$. The uniform model of the 1D CDW insulator \cite{rice,maki} 
has been 
developed earlier and addressed to the macroscopic quasi 1D organics 
and $NbSe_3$ where the quantum tunneling motion has been observed. To
observe FC, the contacts to leads play essential
role. Chamon and Fradkin \cite{cham} argued that the equilibration
along the wide contacts is necessary to have the fractional 
quantized conductance in FQHL. It does not occur in the quantum wire through point-like contacts. Then the fractional
charge does not appear in conductance \cite{pon} or shot noise \cite{2}
in the low energy limit. However, the absence of the equilibration 
gives rise to rich crossovers, where FC manifests itself in many ways.

In this paper we consider transport through a finite length $L$ wire
connected to the source and drain Fermi liquid reservoirs, when the
1D electrons of the wire are in the Mott insulating phase characterized
by the gap $2 M \gg v_c/L \equiv T_L$ in the energy spectrum 
($v_c$: charge velocity in the wire). In spinless case
we describe the transport with an inhomogeneous sin-Gordon action 
after implementing
bosonization and specification of the parameters \cite{shankar}
from comparison with the Bethe-anzats
solution \cite{ktrg} known for the uniform sin-Gordon model. 
For low energy ($<M$),
solution of the inhomogeneous model  
may be done with the quasiclassics \cite{dashen} transformed into 
the instanton technique \cite{pol}. Up to a quantum renormalization
of the soliton mass $M$ this procedure is known to be exact.
The (anti)solitons in this model see the electron reservoirs 
as a slowly decaying interaction along the boundaries 
whose strength is proportional to $q^2$ , 
meanwhile the electrons see the CDW
condensate as quantization of their phase at the boundaries
whose values $2 \pi q \times integer$ relate to the CDW vacua. 
At energies less than $T_L$ the sweeps of the solitons switching 
the CDW phase are deemed instant. Then the problem reduces through the 
Duality Transform \cite{schmid} to the point scatterer
in the TLL whose solution has been known after \cite{kf,fend}.
 Behavior of the low temperature
conductance is ruled by the scaling dimension of the tunneling
operator. The latter is always relevant for the correlated insulator 
opposed to the band insulator where it is marginal. 
Hence the correlated insulator conductance recovers its
free electron value in the absence of impurities passing through
a crossover at the temperature 
$T_x=cst \sqrt{T_L M}exp[-\varepsilon /(1-g)],
\ \varepsilon=\sqrt{M^2-(q \mu)^2}/T_L, \ g=q^2\ $, 
($\mu$: chemical potential). 
Similar, the current
increases linearly with increase of the voltage below $T_x$ and then goes 
down as a negative
power of the voltage.
The tunneling scaling dimension $g$ is related to the quantization of  
the condensate charge which is always fractional except for the case
of the band insulator. 
This FC turns up
in the shot noise of the current at energies above $T_x$. Below $T_x$,
however, the Fermi liquid behavior occurs \cite{2,prb}. 
This
description of the CDW condensate is also applicable to the
charge mode of electrons with spin at rational fillings of even 
denominator $\nu=1/(2m')\times integer$. Therefore, above results
may be addressed to transport through 
the Mott insulator after redefinitions: $q=1/m', \ g=q^2/2$.
In particular, for the Mott-Hubbard insulator $\nu=1/2$ 
the charge $q$ becomes
integer. This marginality shows up in saturation
of the current as voltage exceeds $T_x$ in agreement with \cite{1}.

A general Hamiltonian for spinless electrons moving
in the periodic potential $V_{period}(x)$
( period $a$)
of a one channel wire $0<x<L$
may be written after \cite{pon}
%\begin{equation}
%{\cal H}= \int^{\infty}_{-\infty} d\!x \{ \psi^+(x) ( \epsilon(
% -i \partial_x ) - E_F)\psi + \varphi(x) \rho^2(x) +
%V_{period}(x) \rho(x) \} 
%\label{1}
%\end{equation}
\begin{eqnarray}
\lefteqn{ {\cal H}= \int^{\infty}_{-\infty} d\!x 
\{  \psi^+(x) ( - \frac{
\partial^2_x}{2m^*} - E_F)\psi (x)}
\nonumber\\
& & \hspace{10mm} + u_0 \varphi(x) \rho^2(x) +
V_{period}(x) \rho(x) \} 
\label{1}
%\end{equation}
\end{eqnarray}
where the wire is assumed to be 
smoothly connected to the right and left reservoirs.
The Fermi momentum $k_F$ and the Fermi energy $ E_F$
is determined by the filling factor $\nu$ as $\nu=k_F a/\pi$ 
and $ E_F \approx v_F k_F$.
In Eq. (\ref{1}) the function $\varphi(x)=\theta (x) \theta (L-x)$ 
switches on the
electron-electron interaction inside the wire.
 Following Haldane's 
generalized bosonization procedure \cite{hald} to account for the nonlinear 
dispersion one has to write the fermionic fields as 
$\psi(x)= \sqrt{k_F/(2 \pi)} 
\sum exp\{i(n+1)(k_Fx+\phi(x)/2) +i\theta(x)/2 \}$
and the electron density as $\rho(x)=
(\partial_x \phi(x)+2 k_F)/(2 \pi) 
\sum exp\{in(k_Fx+\phi(x)/2)\}$
where summation runs over even $n$ and 
$\phi, \theta $ are mutually conjugated bosonic fields 
$[\phi(x), \theta(y)]=i 2 \pi sgn(x-y)$. 

After substitution of these expressions into (\ref{1}) the Hamiltonian
takes its bose-form. The associated Eucledian action with this Hamiltonian
can be constructed 
considering the spatial derivative of the $\theta$ field
as a momentum to $\phi$. Its Lagrangian density reads:
\endmulticols
\vspace{-6mm}\noindent\underline{\hspace{87mm}}
\begin{equation}
{\cal L}= \frac{v(x)}{2 g(x)}
\{ { 1 \over {v^2} } 
\left({{\partial_\tau \phi(\tau,x) } \over {\sqrt{4 \pi}}} \right)^2 + 
\left({{\partial_x \phi(\tau,x) } \over {\sqrt{4 \pi}}}
\right)^2 \} -
\frac{E_F^2 U_m}{2 \pi v_F} \varphi(x)
\cos(m[\phi(\tau,x)+2 k_{mF}x])
\label{2}
\end{equation}
\noindent\hspace{92mm}\underline{\hspace{87mm}}\vspace{-3mm}
\multicols{2}\noindent
where we assumed that deviation of the chemical 
potential from that one ($\mu =0$) corresponding to 
$\nu=1/m \times integer$ is small and only one
Umklapp process involving $m$ electron scattering of $2mk_{mF}$ momentum 
transfer is relevant. Here $v_c$ is the charge velocity different
from the Fermi one because of the interaction inside the wire:
for $x \in [0,L]$ the strength of the forward scattering $g(x)=g,\ 0<g<1$
and $v_c=v_F/g$. Outside the wire $g(x)=1$ and $v(x)=v_F$. 
The strengths of the forward $g$ and Umklapp $U_m$ scattering inside 
the wire may be related to the initial parameter $u_0$ in Eq. (\ref{1}). 
When the Umklapp interaction opens up the gap in the spectrum
the parameter $k_{mF}$ does not mean the momentum transfer as in the 
perturbative case. In general, it specifies the value of the chemical
potential as $\mu=k_{mF} v_c/g$.

In the uniform case of the 
infinite wire above action coincides with the sin-Gordon one, which properties 
are known from the Renormalization Group (RG) procedure \cite{ktrg}
extended with the Bethe ansatz solution beyond the perturbation theory. 
The velocity $v$ fixes symmetry between $x$ and $v\tau$ and does not
renormalize with decreasing the energy scale. There are two standard 
characteristics: $K=g m^2-2, \ W=|U_m v_c/v_F|$ of the RG-flow
at zero $k_{mF}$. A hidden
symmetry fixes two separatrices $W=\pm K$. The line $K=-1$ relates to
fermionization \cite{col}. Then the intersection $ W=-K=1$ gives the
point which attracts all trajectories of the massive phase.
These trajectories cover part of the upper half-plane of the parameters
where $K>-2$ except for $K>W$ where $W$ renormalizes
into zero and the TLL phase occurs. Introduction of a non-zero $k_{mF}$ is
not important for the TLL phase. In the massive phase 
the chemical potential for the sin-Gordon elementary excitations, 
solitons destroys the gap on exceeding the soliton mass. Comparison of the 
parameters of (\ref{2}) with the Bethe-anzats 
solution shows that the local interaction
producing a gap has to be strong enough for the spinless electron
system. In the quantum wire such a local interaction 
occurs if the wire width
$w$ is about the distance $d$ from the wire to the screening gate.
In the opposite case of the long range interaction the initial value
of the constant $g \propto 1/\sqrt{d/w}$ is small \cite{schulz} and 
the uniform sin-Gordon model (\ref{2}) with $m$ larger than 2
may acquire the gap at appropriate rational filling. 
The same sin-Gordon action (\ref{2}) describes also
the \emph{even } Umklapp processes in the 1D system of the interacting 
electrons  with spin \cite{us}, if the $\phi$ field denotes 
the charge bose field, $m$ under the $cos$ in (\ref{2}) is
changed into $m/\sqrt{2}$ and $k_{mF}$ into $\sqrt{2} k_{mF}$. 
Contrary to the spinless case, any repulsive interaction will 
open a gap in the spectrum of the 1D spin electron system at half-filling 
$m=2$ since $g=1$ is critical:$\ K=0$. 
Carrying out all consideration below
for the spinless electrons, we will notice a necessary modification
in the spin case at the end.

\emph{Duality Transform } - 
In the massive phase of the  wire where $\mu$
is much less than the mass $M$ of the solitons the quasiclassical method
gives correct physical picture \cite{dashen} up to a quantum 
renormalization of the parameters. Then tunneling of the charge through
the wire may be described in the instanton technique \cite{pol}. 
First, let us integrate out the field $\phi$ in the reservoirs 
$x \bar{\in} [0,L]$ in (\ref{2}). It can be done, say, for $x \le 0$ 
with introducing a new variable: 
$\xi(\tau,x)=\phi(\tau,x)-\phi(\tau,0) e^{x/L'}$ satisfying the boundary
condition $\xi(\tau,0)=0$. Here parameter
$L' \rightarrow \infty$ regularizes the infinite length of the left
reservoir. Substitution of the new variable into (\ref{2}) brings up
a Gaussian action for $\xi$, which after integration produces the 
boundary action describing non-local interaction for $\phi(\tau,0)$ . 
Repeating the same procedure
for $x \ge L$ we can write this boundary action as
\begin{equation}
{\cal S}_{int}=\frac{Re}{16 \pi ^2} \sum_{y=0,L}
\int \int d \tau d \tau' \left(\frac{\phi(\tau,y)-\phi(\tau',y)}
{\tau - \tau' -i \alpha} \right)^2
\label{4}
\end{equation}
To simplify notations we will drop out $v_c$ below. 
Then all energies become
transformed into reverse lengths in line with: 
$length^{-1} = energy / v_c$. It assumes, in particular,
rescaling $\tau$ in (\ref{4}) so that a new one 
is previous $v_c \tau$, and the cut-off $\alpha$ becomes $\alpha=v_c/E_F$.
The rest part of the action (\ref{2}) may be re-written as 
\endmulticols
\vspace{-6mm}\noindent\underline{\hspace{87mm}}
\begin{equation}
{\cal S}_{blk}= \int d\tau \{ \frac{1}{4 \pi g m^2}\int^L_0 d x 
\frac{m^2}{2}
[( \partial_\tau \phi(\tau,x))^2 + ( \partial_x \phi(\tau,x))^2 ] +
\lambda^2(1- \cos(m\phi(\tau,x))) -
\frac{k_{mF}}{2 g \pi}[\phi(\tau,L)-\phi(\tau,0)] \}
\label{5}
\end{equation}
\noindent\hspace{92mm}\underline{\hspace{87mm}}\vspace{-3mm}
\multicols{2}\noindent
In this "bulk" part of the action where $\lambda^2=2g W m^2/\alpha^2$
and the definition of 
the $\phi$ field has been changed with respect to (\ref{2}) to
absorb $2k_{mF}x$ function, the $cos$ term tends to fix this new
field $\phi$ at one of the $integer \times 2 \pi/m$ values. These 
values correspond to the infinite set of degenerate vacua of the
infinite length sin-Gordon model. At finite length, however, the
non-degenerate ground state is recovered due to an exponentially 
weak tunneling between the vacua. The tunneling may  be conventionally
described in the instanton technique \cite{pol}. In this technique the 
partition function: 
${\cal Z}=\int D\phi exp[-{\cal S}_{blk}]$ 
that will be considered first for the action (\ref{5})
omitting the boundary interaction is calculated in an "advanced"
saddle point approximation. The extremum 
function, (anti-)instanton, minimizing the action (\ref{5})
coincides with the sin-Gordon stationary 
(anti)soliton and takes the form:
$\pm \phi_0(\tau,x|\tau_0)=m^{-1}
f(\lambda [(x-L/2)\cos \varpi -(\tau-\tau_0)\sin \varpi]),\ 
f(x)=4 arctan(exp(x))$. The angle $\varpi \in [0,\pi )$ 
satisfies:
\begin{eqnarray}
{\cal S}_{blk}\{\pm \phi_0 \}= \frac{L}{\sin\varpi }
\left(M \pm \frac{k_{mF} \cos\varpi }{g m} \right)\equiv L s_0
\label{6} \\
\partial_{\varpi} s_0=0 \ : \ \left\{ \matrix{
\cos \varpi =\mp k_{mF}/(g m M)=\mp q \mu /M \cr
s_0=M \sin \varpi =\sqrt{M^2-(q\mu )^2} \cr } \right.
\label{7}
\end{eqnarray}
being different for the instanton and anti-instanton.
Here the mean field value of $M$ is a classical 
(anti)soliton mass $M_{cl}=2 \lambda/(\pi m^2 g)$ and 
the minimum of the action ${\cal S}_{blk}$
exists while $M > k_{mF}/(g m)=q \mu$.
The instanton/
anti-instanton varies the phase $\phi$ by $\mp 2 \pi/m$ and carries 
the charge $\pm q,\ q=1/m$.
Both limit values of $\phi $ are approached exponentially as $\tau $
moving away from the kink location $\tau_0$. Therefore, a many 
(anti-)instanton function compiled from a number of kinks 
which locations in imaginary time $\tau_j$ are well separated from 
each other, presents an approximate minimum function for the action 
(\ref{5}). The partition function reduces to a sum of 
the contributions of all these minima.  The contribution to 
the functional integral of ${\cal Z}$ may be found
expanding the action around each (anti-)instanton 
($e_j=\pm $) of the many kink function 
to the second order and considering each kink independently
\begin{equation}
{\cal Z}=cst(1+\sum_{N=1}^\infty 1/N! \int 
\prod_{j=1}^N \{d\tau_j \sum_{e_j=\pm} p_{e_j} e^{-Ls_0}\} )
\label{8}
\end{equation}
Here the $cst$ has absorbed the infinite sum over all degenerate vacua. 
The prefactors $p_{\pm}$ for 
the (anti-)instanton are equal and may be calculated as
\begin{eqnarray}
p_{\pm}&=\sqrt{\int dx \int d \tau \frac{[\phi_0'(\tau,x)]^2}{4 \pi g m^2}} 
\left\{\frac{Det(-\partial^2+\lambda^2)}
{Det'(-\partial^2+\lambda^2 \cos\phi_0)}\right\}^{1/2}
\nonumber \\
&=\sqrt{\frac{M \sin \varpi}{m^2 L}}e^{-L \Delta M/\sin \varpi}
\label{9}
\end{eqnarray}
Here the determinant $Det$ of the first operator $-\partial^2+\lambda^2 \equiv 
-\partial_\tau^2-\partial_x^2+\lambda^2 $ has to be calculated in the strip. 
The same has to be done for determinant $Det'$. 
However, the latter does not include
contribution of the zero energy mode
of the second Shr\"{o}dinger operator $-\partial^2+\lambda^2 \cos\phi_0$.
This operator differs with the first one due to 
an additional potential well produced by the instanton. Equality of
the lowest eigenvalue of  the second operator to zero ensues from
the translational invariance of the action (\ref{5}) in time.  
The result of calculation of (\ref{9}) $p$ has an exponent with  
the one loop quantum correction to the  classical soliton mass 
\cite{dashen}: $\Delta M= \sum \Delta \omega /2$ multiplied by $L$
where $\Delta \omega$ is difference between the eigenvalues of the operators
depending on $\tau$ only and constructed from the first 
and second one, respectively, by dropping out their $x$ dependence. 
This correction renormalizes the value of the mass $M=M_{cl}$ in (\ref{6}) 
found earlier and, hence, the value of the angle $\varpi$ in (\ref{7}). 
One can see that taking into account the higher order 
corrections amounts to the complete RG-procedure for the uniform sin-Gordon action 
which ends up with $M$ equal to the quantum soliton mass and the renormalized
value of $g$: $g \rightarrow 1/m^2$ in above equations.  

Incorporation of the boundary interaction (\ref{4}) in above consideration
entails following changes
in the instanton representation of the partition function (\ref{8}). 
Besides a dimensionless factor multiplying the ratio of the determinants
due to their boundary variation, there
appears an additional contribution to the one kink mean field action 
together with the direct interaction between the kinks. 
The infrared divergency in these terms limits the sum over $e_j=\pm$ 
to the neutral combinations: $\sum_j e_j =0$, for which the boundary
part of the exponent becomes 
${\cal S}_{int}={-1 \over m^2} [\sum_{i \neq j}^N e_i e_j 
F_{e_i e_j}(\tau_i - \tau_j)-N ln(M\sin\varpi )]$. 
Both interactions are equal 
$F_\pm(\tau )=ln \sqrt{(\tau_i - \tau_j )^2+1/D'^2}$
at long time $\tau > 1/D'$, but differ at short.  
At $\sin \varpi =1$ the time $1/D'$ comes about through 
bending the straight linear
form of the kinks on their approaching each other. It can be
evaluated as $D'=\sqrt{T_L M}$. For $q \mu \gg \sqrt{T_L M}$
the time is approximately an 
extension of the single kink $L/tan \varpi $ and  
$D'=T_L \sqrt{(M/q \mu )^2-1} \approx T_L M/(q \mu)$.
Integrating out the energies $> D'$ in (\ref{8}) results in 
renormalization of the prefactor $P(D')$. In the lowest order
in exponentially small kink amplitude only 
the kink-antikink pairs contribute. The renormalized
prefactor is
\[
P=p \times e^{-{1 \over m^2}\left[2 \int_{D'}^{M \sin\varpi}
F_-(\omega) d \omega -F_-(\tau =0) -ln(M \sin\varpi) \right] }
\]
If at $\sin \varpi =1$ the function $F_-(\tau )$ quickly drops
reaching  the bottom value $F_-(\tau )=-lnM$ at 
$\tau < 1/D'$, this equation yields
$P \simeq p = cst D'$, which 
complies with the exact solution $P= D'\sin \varpi/ \pi$ 
known for the band
insulator $m=1$.  On the other hand, assuming correctness of 
the Gaussian approximation, the renormalized prefactor 
may be found in general without specification of $F$ in above
scheme as $P=cst \times 
D'\sin \varpi (\sqrt{T_L M/ \sin \varpi}/D')^{1-1/m^2}$. 

\emph{Low energy transport } - 
Constructed Dual Transform of the low energy partition function may be 
identified with the partition function
for the point scatterer in the TLL \cite{kf} in the following way.  
We introduce a bosonic field $\theta(\tau,x)$ after  Schmid \cite{schmid}
and will ascribe to each (anti-)instanton in (\ref{8}) a factor 
$exp\{\mp \theta(\tau_j,0)/m \}$. 
 Having a small voltage $V$ been applied, the real time Lagrangian
for the $\theta$ field reads 
\begin{eqnarray}
\int d x {\cal L}_t&=& \int {dx \over 2}  
\{ 
\left({{\partial_t \theta(t,x) }
     \over {\sqrt{4 \pi}}}
\right)^2 -
\left({{\partial_x \theta(t,x) } \over {\sqrt{4 \pi}}} \right)^2 
\}
\nonumber \\
&-& {D'A \over \pi } \cos((V t/v_c + \theta(t,0))/ m) 
\label{10}
\end{eqnarray}
where $t$ has a length dimension and 
the amplitude of the fractional charge tunneling $A=2 \pi P
e^{-L M \sin \varpi }/D'$ is exponentially small.  The 
tunneling current comes about through varying 
the tunneling part of the Hamiltonian
associated to (\ref{10}):
$J={\delta \over \delta \theta } {\cal H }_{tunn}=-A D' \sin((\theta + V t)/m)/( \pi m) $.
On rescaling $\tilde{\theta }= \theta /m $, the tunneling may be regarded as a 
point scatterer of 
the integer charge in the uniform TLL with $g=m^{-2}=q^2$. Solution
of this problem first considered asymptotically \cite{kf} has been later
thoroughly examined in the Bethe anzats technique \cite{fend}. In this
problem the current $J$
can be identified as  the $m$ times backscattering current 
under applied $mV$ voltage:
both the backscattering current and the effective voltage
are proportional to $g$ \cite{2} in the uniform TLL. 
Its zero temperature
expression takes a closed form:
\begin{equation}
{J \over V \sigma_0}={i  \over \sqrt{\pi } } \int_C {dz \over z} 
\frac{\Gamma(1+{z \over 1-g})
\Gamma(1+{z \over 1-1/g}) }{4 \Gamma(3/2 +z)} 
\left({mV \over T_x} \right)^{2z}
\label{11}
\end{equation}
where the contour $C$ goes along the imaginary axis circumventing 
the zero from the left, $\sigma_0$ is equal to $1/(2 \pi )$ 
in spinless case
and $T_x={2 \over g} D' ( A/\Gamma(g))^{1/(1-g)}$ \cite{weiss} 
gives a crossover
energy. Substituting the value of $A$, one can find 
$T_x \simeq 
cst \sqrt{T_L M} e^{-M \sin\varpi /[(1-g)T_L]}$. 
Below $T_x$ the conductance $J/V$ approaches its universal
value $\sigma_0$ as $\sigma_0(1 - c_1(1/g) \times (m V/T_x)^{2(1/g-1)})$ 
 and $c_1(g)=\sqrt{\pi } \Gamma(g+1)/(2 \Gamma(1/2+g))$. 
Above $T_x$ it is going to zero as $\sigma_0 c_1(g) (m V/ T_x)^{2(g-1)}$ 
reaching an exponentially small value at $V \approx T_L$.
The linear bias conductance vs. temperature relates to above voltage 
dependence through exchange $T \leftrightarrow V$ 
with the same exponents in the low and high
temperature asymptotics. 
In the band insulator case $m=1$ the crossover temperature $T_x$ is zero 
and the conductance 
reduces to the constant: 
$\sigma_0 A^2$. 
For larger values of $m$, however,
it becomes non-monotonic function of $V$ with a maximum at $V \approx T_x$.

Next, let us examine the shot noise of the current $\delta J^2$. 
Since at $T=0$ it 
coincides with the zero frequency noise of the backscattering current
\cite{2}, we may read off the result from the shot noise of 
the point scatterer in the TLL
\cite{noise,weiss} as: $
\delta J^2(V)=-\frac{mV^2}{2(1/g-1)} \partial_V {J \over V}$.
In the high voltage $mV > T_x$ regime 
this equation reveals that the charge of the transport carrier is
fractional $q$: $\delta J^2=q J$, while for the low voltage
the equation takes its Fermi liquid form: $\delta J^2=m(\sigma_0 V-J)$. The
factor $m$ in the latter case shows that the backscattering is
the $m$-electron process. Addition of a weak impurity scattering changes
the leading low energy asymtotics into the $m=1$ form. However, the 
high voltage form remains the same 
since the instanton representation of tunneling between the CDW vacua
holds on in the presence of a weak one electron
backscattering after a slight modification of the (anti-)instanton mass
and the tunneling amplitude.  
The last speculation concurs with the earlier considerations \cite{rice,maki}
of the infinite Peierls CDW transport, and can be confirmed with the 
exact  solution for 1D transport through the Mott-Hubbard insulator with
impurity \cite{3}. 

Modification of above consideration for the spin case of filling factor
with even denominator $\nu=1/m=1/(2m')$
ensues from above correspondence between spinless and spinful cases:
$m \rightarrow m/\sqrt{2}; k_{mF} \rightarrow \sqrt{2}k_{mF}$ and has to 
account for an additional factor $\sqrt{2}$ in relation between
the charge density and the fields $\phi, \theta $ variation. 
Therefore, $q=2/m, g=q^2/2$ and $k_{mF}/m$ transforms into $q \mu$ as before
for the definition of the chemical potential remains the same.
Voltage $V$ in (\ref{10}) acquires the factor $\sqrt{2}$ which together
with another $\sqrt{2}$ after rescaling $m$ 
gives correct combination $qV$ under $\cos$.
All equations below (\ref{10}) have no change in their
form assuming that above parameters are modified and 
$\sigma_0$ becomes $1/\pi $. So, there is no change in
the shot noise expressions.  In spite of the similar form,
the results for the Mott insulator have more 
complicated physics behind, as the reservoir electrons now transform into
two types of quasiparticles carrying either spin or charge. 
In particular, the integer charge ($q=1$) of the charge quasiparticle 
of the Mott-Hubbard insulator $m'=1$ does not assume 
absence of the non-trivial crossover \cite{1} anymore.

Finally, the constructed low energy solution brings up a new insight on
the earlier developed theory of transport in the infinite 1D CDW \cite{rice,maki}. 
It shows that as the conductance tends to vanish at 
the energies $\approx T_L$, the CDW phase becomes fixed by 
the reservoirs. The latter has been 
a crucial assumption for deriving the thermally activated conductivity
in the CDW materials. On the other hand, recovering of the 
conductance below $T_x$ calls up an earlier Fr\"{o}lich's idea of the CDW
sliding in the absence of impurities \cite{lee}.

The authors acknowledge discussion with A. Odintsov.
This work has been supported by the Center of Excellence at the JSPS.

\end{document}